\documentclass[a4paper,12pt,draftcls,technote,onecolumn]{IEEEtran}%
\usepackage{mathrsfs}
\usepackage{amsfonts}
\usepackage{amsmath}
\usepackage{amssymb}
\usepackage{graphicx}%

\ifCLASSINFOpdf
\else
\fi
\begin{document}
%
\title{Additive-State-Decomposition-Based Output Feedback Tracking Control for
Systems with Measurable Nonlinearities and Unknown Disturbances}
%
%
%

\author{Quan Quan, Kai-Yuan Cai, Hai Lin
\thanks{Quan Quan is with Department of Automatic Control, Beihang University, Beijing, 100191,
China, was with Department of Electrical and Computer Engineering,
National University of Singapore, Singapore, 117576, Singapore. Email: (see http://quanquan.buaa.edu.cn).}
\thanks{Kai-Yuan Cai  is with Department of Automatic Control, Beihang University, Beijing, 100191,
China.}
\thanks{Hai Lin is with Department of Electrical and Computer Engineering,
National University of Singapore, Singapore, 117576, Singapore.}
}

%
%

\markboth{}%
{Shell \MakeLowercase{\textit{et al.}}: Bare Demo of IEEEtran.cls for Journals}
%



\maketitle

\begin{abstract}
In this paper, a new control scheme, called
\emph{additive-state-decomposition-based tracking control}, is
proposed to solve the output feedback tracking problem for a class
of systems with measurable nonlinearities and unknown disturbances.
By the additive state decomposition, the output feedback tracking
task for the considered nonlinear system is decomposed into three
independent subtasks: a pure tracking subtask for a linear time
invariant (LTI) system, a pure rejection subtask for another LTI
system and a stabilization subtask for a nonlinear system. By
benefiting from the decomposition, the proposed
additive-state-decomposition-based tracking control scheme i) can
give a potential way to avoid conflict among tracking performance,
rejection performance and robustness, and ii) can mix both designs
in time domain and frequency domain for one controller design. To
demonstrate the effectiveness, the output feedback tracking problem
for a single-link robot arm subject to a sinusoidal or a general
disturbance is solved respectively, where the transfer function
method for tracking and rejection and the feedback linearization
method for stabilization are applied together to the design.
\end{abstract}


%
\IEEEpeerreviewmaketitle

\section{Introduction}

In this paper, the output feedback tracking problem for a class of
systems with measurable nonlinearities and unknown disturbances is
considered. This problem has attracted great research interest in
recent years \cite{Ding(2003)}-\cite{Bobtsov(2011)}. As far as
nonlinear systems are concerned, several results are available under
the minimum phase assumption. In \cite{Ding(2003)}, global
disturbance rejection with stabilization for nonlinear systems in
output feedback form was solved in spite of the disturbance
generated by a finite dimensional exosystem. The similar problem but
subject to unknown parameters on both input matrix and system matrix
was considered in \cite{Riccardo(2005a)}. In \cite{Bobtsov(2011)},
another control algorithm in simplicity of implementation was
proposed for nonlinear plants with parametric and functional
uncertainty in the presence of biased harmonic disturbance. The
problem about nonminimum phase nonlinear systems was further
considered. In \cite{Riccardo(2005b)}-\cite{Ding(2006)}, adaptive
estimation of unknown disturbances in a class of nonminimum phase
nonlinear systems, and the stabilization and disturbance rejection
based on the estimated disturbances for single-input single-output
(SISO) systems were solved. The result was further extended to a
class of nonminimum phase nonlinear Multiple-Input Multiple-Output
(MIMO) systems in \cite{Lan(2006)}. In \cite{Riccardo(2008)}, a
solution to this problem was provided for nonminimum phase systems
with uncertainties in both parameters and order of an exosystem.

The basic idea of the current work is to decompose the output
feedback tracking task\ into simpler subtasks. Then one can design a
controller for each subtask respectively, which are finally
integrated together to achieve the original control task. The
motivation of this paper can be described as follows. First, it is
to avoid conflict among tracking performance, rejection performance
and robustness. It is well known that there is an intrinsic conflict
between performance (trajectory tracking and disturbance rejection)
and robustness in the standard feedback framework \cite{Chen(1995)}%
,\cite{Zhou(1996)}. By the control scheme mentioned in \cite{Ding(2003)}%
-\cite{Bobtsov(2011)}, as the dimension of the exosystem is
increasing, the closed-loop system will be incorporated into a copy
of marginally stable exosystem according to internal model principle
\cite{Francis(1976)} to achieve high performance (trajectory
tracking and disturbance rejection). The price to be paid is a
reduced robustness against uncertainties, especially for nonminimum
phase systems. Moreover, conflict between tracking performance and
rejection performance exists as well when reference and disturbance
behave differently \cite{Morari(1989)}. Secondly, it is to relax the
restriction on
the disturbance. For the control scheme mentioned in \cite{Ding(2003)}%
-\cite{Riccardo(2008)}, a \textquotedblleft regulator
equation\textquotedblright\ often needs to be solved first for a
coordinate transformation, which yields an error system with
disturbances appearing at the input. However, these control schemes
are only applicable to finite-dimensional autonomous exosystems.
While, the control scheme mentioned in \cite{Bobtsov(2011)} requires
the system being minimum phase to shift disturbances to the input
channel.

For such a purpose, a new control scheme based on the \emph{additive
state decomposition}\footnote{In this paper we have replaced the
term \textquotedblleft additive decomposition\textquotedblright\ in
\cite{Quan(2009)} with the more descriptive term \textquotedblleft
additive state decomposition\textquotedblright.}, called
\emph{additive-state-decomposition-based tracking control}, is
proposed which is applicable to both minimum phase and nonminimum
phase systems. The proposed additive state decomposition is a new
decomposition manner different from the lower-order subsystem
decomposition methods existing in the literature, see e.g.
\cite{Fradkov(1999)},\cite{Zhu(2010)}. Concretely, taking the system
$\dot{x}\left(  t\right)  =f\left(  t,x\right)  ,x\in%
\mathbb{R}
^{n}$ for example, it is decomposed into two subsystems:
$\dot{x}_{1}\left( t\right)  =f_{1}\left(  t,x_{1},x_{2}\right)  $
and $\dot{x}_{2}\left(
t\right)  =f_{2}\left(  t,x_{1},x_{2}\right)  $, where $x_{1}\in%
\mathbb{R}
^{n_{1}}\ $and$\ x_{2}\in%
\mathbb{R}
^{n_{2}},$ respectively. The lower-order subsystem decomposition satisfies%
\[
n=n_{1}+n_{2}\text{ and }x=x_{1}\oplus x_{2}.
\]
By contrast, the proposed additive state decomposition satisfies%
\[
n=n_{1}=n_{2}\text{ and}\ x=x_{1}+x_{2}.
\]
In our opinion, lower-order subsystem decomposition aims to reduce
the complexity of the system itself, while the additive state
decomposition emphasizes the reduction of the complexity of tasks
for the system.

By following the philosophy above, in the
additive-state-decomposition-based tracking\emph{\ }control scheme,
the output feedback tracking is `additively' decomposed into three
independent subtasks, namely the tracking subtask, the rejection
subtask and the stabilization subtask. Three subcontrollers for the
three subtasks are designed separately then. Since the resulting
controller possesses three degrees of freedom, the proposed scheme
in fact gives a potential way to avoid conflict among tracking
performance, rejection performance and robustness. Moreover, by the
additive-state-decomposition-based tracking\emph{\ }control scheme,
it will be seen that both the tracking subtask and rejection subtask
only need to be achieved on a linear time invariant (LTI) system.
Consequently, the tracking controller and disturbance compensator
can be designed in both time domain and frequency domain. In this
framework, the existing output regulation methods as in
\cite{Ding(2003)}-\cite{Riccardo(2008)} can be incorporated. Also,
it can take advantage of some standard design methods in frequency
domain to handle general disturbances. More importantly, nonminimum
phase systems can be handled in the same framework.

This paper is organized as follows. In Section 2, the problem
formulation is given and the additive state decomposition is
recalled briefly first. In Section 3, the considered system is
transformed to a disturbance-free system in sense of input-output
equivalence. Sequently, in Section 4, the transformed system is
`additively' decomposed into three subsystems. In Section 5,
controller design is given. Section 6 concludes this paper.

\section{Problem Formulation and Additive Decomposition}

\subsection{Problem Formulation}

Consider a class of SISO nonlinear systems similar to \cite{Ding(2003)}%
-\cite{Ding(2006)},\cite{Riccardo(2008)}-\cite{Bobtsov(2011)}:%
\begin{align}
\dot{x}  &  =A_{0}x+bu+\phi_{0}\left(  y\right)  +d,x\left(
0\right)
=x_{0}\nonumber\\
y  &  =c^{T}x \label{equ0}%
\end{align}
where $A_{0}\in%
\mathbb{R}
^{n\times n}\ $is a constant matrix, $b\in%
\mathbb{R}
^{n}\ $and $c\in%
\mathbb{R}
^{n}\ $are constant vectors, $\phi_{0}:%
\mathbb{R}
\rightarrow%
\mathbb{R}
^{n}$ is a nonlinear function vector, $x\left(  t\right)  \in%
\mathbb{R}
^{n}$ is the state vector, $y\left(  t\right)  \in%
\mathbb{R}
$ is the output, $u\left(  t\right)  \in%
\mathbb{R}
$ is the control, and $d\left(  t\right)  \in%
\mathbb{R}
^{n}$ is a bounded disturbance. It is assumed that only $y$ is
available from
measurement. The desired trajectory $r\left(  t\right)  \in%
\mathbb{R}
$ is known and smooth enough, $t\geq0$. In the following, for
convenience, we will omit the variable $t$ except when necessary.

\textbf{Remark 1}\textit{.} Under certain conditions, the system in the form%
\begin{align*}
\dot{x}  &  =f\left(  x\right)  +g\left(  x\right)  u+d\\
y  &  =h\left(  x\right)
\end{align*}
can be transformed to (\ref{equ0}). The sufficient and necessary
condition to ensure the existence of transformation can be found in
\cite{Marino(1993)}.

\textbf{Remark 2}\textit{. }The considered\textit{\ }SISO nonlinear
system (\ref{equ0}) is allowed to be a nonminimum phase\textit{\
}system, i.e., the transfer function of the linear part (i.e.,
regardless of nonlinear dynamics
$\phi_{0}\left(  y\right)  $ and disturbance $d$)%
\[
c^{T}\left(  sI-A_{0}\right)  ^{-1}b=\frac{N\left(  s\right)
}{D\left(
s\right)  }%
\]
is nonminimum phase here, where $N\left(  s\right)  $ has zeros on
the right $s$-plane. It is noticed that the property of nonminimum
phase cannot be changed by output feedback.

For system (\ref{equ0}), the following assumption is made.

\textbf{Assumption 1}. The pair $\left(  A_{0},c\right)  $ is
observable.

Under \textit{Assumption 1},\textit{\ }the objective here is to
design a tracking controller $u$ such that $y\rightarrow r$ as
$t\rightarrow\infty\ $or with good tracking accuracy, i.e, $y-r$ is
ultimately bounded by a small value.

\subsection{Additive State Decomposition}

In order to make the paper self-contained, additive state
decomposition \cite{Quan(2009)} is recalled here briefly. Consider
the following `original'
system:%
\begin{equation}
f\left(  {t,\dot{x},x}\right)  =0,x\left(  0\right)  =x_{0}
\label{Gen_Orig_Sys}%
\end{equation}
where $x\in%
\mathbb{R}
^{n}$. We first bring in a `primary' system having the same
dimension as
(\ref{Gen_Orig_Sys}), according to:%
\begin{equation}
f_{p}\left(  {t,\dot{x}_{p},x_{p}}\right)  =0,x{_{p}}\left(
0\right)
=x_{p,0} \label{Gen_Pri_Sys}%
\end{equation}
where ${x_{p}}\in%
\mathbb{R}
^{n}$. From the original system (\ref{Gen_Orig_Sys}) and the primary
system
(\ref{Gen_Pri_Sys}) we derive the following `secondary' system:%
\begin{equation}
f\left(  {t,\dot{x},x}\right)  -f_{p}\left(
{t,\dot{x}_{p},x_{p}}\right)
=0,x\left(  0\right)  =x_{0} \label{Gen_Sec_Sys0}%
\end{equation}
where ${x_{p}}\in%
\mathbb{R}
^{n}$ is given by the primary system (\ref{Gen_Pri_Sys}). Define a
new
variable ${x_{s}}\in%
\mathbb{R}
^{n}$ as follows:%
\begin{equation}
{x_{s}\triangleq x-x_{p}}. \label{Gen_RelationPS}%
\end{equation}
Then the secondary system (\ref{Gen_Sec_Sys0}) can be further
written as
follows:%
\begin{equation}
f\left(  {t,\dot{x}_{s}+\dot{x}_{p},x_{s}+x_{p}}\right)
-f_{p}\left( {t,\dot{x}_{p},x_{p}}\right)  =0,x{_{s}}\left(
0\right)  =x_{0}-x_{p,0}.
\label{Gen_Sec_Sys}%
\end{equation}
From the definition (\ref{Gen_RelationPS}), we have%
\begin{equation}
{x}\left(  t\right)  ={x_{p}\left(  t\right)  +x_{s}\left(  t\right)
,t\geq0.} \label{Gen_RelationPS1}%
\end{equation}

\textbf{Remark 3}\textit{.} By the additive state
decomposition\textit{, }the\textit{\ }system (\ref{Gen_Orig_Sys}) is
decomposed into two subsystems with the same dimension as the
original system. In this sense our decomposition is
\textquotedblleft additive\textquotedblright. In addition, this
decomposition is with respect to state. So, we call it
\textquotedblleft additive state
decomposition\textquotedblright\emph{.}

As a special case of (\ref{Gen_Orig_Sys}), a class of differential
dynamic
systems is considered as follows:%
\begin{align}
\dot{x}  &  =f\left(  {t,x}\right)  ,x\left(  0\right)  =x_{0},\nonumber\\
y  &  =h\left(  {t,x}\right)  \label{Dif_Orig_Sys}%
\end{align}
where ${x}\in%
\mathbb{R}
^{n}$ and $y\in%
\mathbb{R}
^{m}.$ Two systems, denoted by the primary system and (derived)
secondary
system respectively, are defined as follows:%
\begin{align}
\dot{x}_{p}  &  =f_{p}\left(  {t,x_{p}}\right)  ,x_{p}\left(
0\right)
=x_{p,0}\nonumber\\
y_{p}  &  =h_{p}\left(  {t,x}_{p}\right)  \label{Dif_Pri_Sys}%
\end{align}
and%
\begin{align}
\dot{x}_{s}  &  =f\left(  {t,x_{p}}+{x_{s}}\right)  -f_{p}\left(  {t,x_{p}%
}\right)  ,x_{s}\left(  0\right)  =x_{0}-x_{p,0},\nonumber\\
y_{s}  &  =h\left(  {t,x_{p}}+{x_{s}}\right)  -h_{p}\left(
{t,x}_{p}\right)
\label{Dif_Sec_Sys}%
\end{align}
where ${x_{s}}\triangleq{x-x_{p}}$ and $y_{s}\triangleq{y-y_{p}}$.
The secondary system (\ref{Dif_Sec_Sys}) is determined by the
original system (\ref{Dif_Orig_Sys}) and the primary system
(\ref{Dif_Pri_Sys}). From the
definition, we have%
\begin{equation}
{x}\left(  t\right)  ={x_{p}\left(  t\right)  +x_{s}\left(  t\right)
,y\left(  t\right)  =y_{p}\left(  t\right)  +y_{s}\left(  t\right)
,t\geq0.}
\label{Gen_RelationDif}%
\end{equation}

\section{Model Transformation}

Firstly, we need to estimate the state from the output. The main
difficulty is
how to handle the disturbances in the state equation. In \cite{Ding(2003)}%
-\cite{Bobtsov(2011)}, an extended state observer including states
of the considered nonlinear system and the exosystem is designed,
where the disturbance is assumed to be generated by a
finite-dimensional autonomous exosystem. According to this, the
model of the disturbance has in fact determined the performance of
the observation partly. However, in practice, a general disturbance
is difficult to model as a finite-dimensional autonomous one, or
with uncertainties. To tackle this difficulty, we first transform
the original system (\ref{equ0}) to a disturbance-free system, which
is proved to be input-output equivalent with the aid of the additive
state decomposition as stated in the following theorem.

\textbf{Theorem 1}\textit{.} Under \textit{Assumption 1}, there
exists a
vector $p\in%
\mathbb{R}
^{n}$ such that $A=A_{0}+pc^{T}\ $is stable, and the system
(\ref{equ0}) is
input-output equivalent to the following system:%
\begin{align}
\dot{x}_{new}  &  =Ax_{new}+bu+\phi\left(  y\right)  ,x_{new}\left(
0\right)
=0\nonumber\\
y  &  =c^{T}x_{new}+d_{new} \label{equ1_equivalent}%
\end{align}
where\ $\phi\left(  y\right)  =\phi_{0}\left(  y\right)  -py$ and%
\begin{equation}
x_{new}=x-\left(  e^{At}x_{0}+\int\nolimits_{0}^{t}e^{A\left(
t-s\right)
}d\left(  s\right)  ds\right)  ,d_{new}=c^{T}\left(  e^{At}x_{0}%
+\int\nolimits_{0}^{t}e^{A\left(  t-s\right)  }d\left(  s\right)
ds\right)  .
\label{equ1_equivalent_pro}%
\end{equation}

\textit{Proof.} Since the pair $\left(  A_{0},c\right)  $ is
observable, there
always exists a vector $p\in%
\mathbb{R}
^{n}$ such that $A=A_{0}+pc^{T}$ is stable, whose the eigenvalues
can be
assigned freely. The system (\ref{equ0}) can be rewritten as follows:%
\begin{align}
\dot{x}  &  =Ax+bu+\phi\left(  y\right)  +d,x\left(  0\right)  =x_{0}%
\nonumber\\
y  &  =c^{T}x \label{equ1}%
\end{align}
where $\phi\left(  y\right)  =\phi_{0}\left(  y\right)  -py$. In the
following, additive state decomposition is utilized to decompose the
system (\ref{equ1}). Consider the system (\ref{equ1}) as the
original system and
choose the primary system as follows:%
\begin{align}
\dot{x}_{p}  &  =Ax_{p}+d,x_{p}\left(  0\right)  =x_{0}\nonumber\\
y_{p}  &  =c^{T}x_{p}. \label{equ1_Pri0}%
\end{align}
Then the secondary system is determined by the original system
(\ref{equ1}) and the primary system (\ref{equ1_Pri0}) with the rule
(\ref{Dif_Sec_Sys})
that%
\begin{align}
\dot{x}_{s}  &  =Ax_{s}+bu+\phi\left(  y\right)  ,x_{s}\left(
0\right)
=0\nonumber\\
y_{s}  &  =c^{T}x_{s}. \label{equ1_Sec0}%
\end{align}
According to (\ref{Gen_RelationDif}), we have $x=x_{p}+x_{s}\ $and$\ y=y_{p}%
+y_{s}.$ Therefore, the following system is an input-output
equivalent system
of (\ref{equ1}):%
\begin{align}
\dot{x}_{s}  &  =Ax_{s}+bu+\phi\left(  y\right)  ,x_{s}\left(
0\right)
=0\nonumber\\
y  &  =c^{T}x_{s}+c^{T}x_{p} \label{equ1_equivalent_1}%
\end{align}
where $x_{p}$ is generated by (\ref{equ1_Pri0}). By
(\ref{equ1_Pri0}), it holds that
$x_{p}=e^{At}x_{0}+\int\nolimits_{0}^{t}e^{A\left(  t-s\right)
}d\left(  s\right)  ds.$ Let $x_{s}=x_{new}$ and
$d_{new}=c^{T}x_{p}.$ Substituting into (\ref{equ1_equivalent_1})
yields (\ref{equ1_equivalent}). $\square$

For the disturbance-free transformed system (\ref{equ1_equivalent}),
we design an observer to estimate $x_{new}$ and $d_{new}$, which is
stated in \textit{Theorem 2.}

\textbf{Theorem 2}\textit{.} Under \textit{Assumption 1}, an
observer is designed to estimate state $x_{new}$ and $d_{new}$ in
(\ref{equ1_equivalent})
as follows%
\begin{align}
\dot{\hat{x}}_{new}  &  =A\hat{x}_{new}+bu+\phi\left(  y\right)
,\hat
{x}_{new}\left(  0\right)  =0\nonumber\\
\hat{d}_{new}  &  =y-c^{T}\hat{x}_{new}. \label{equ1_equivalent_est}%
\end{align}
Then $\hat{x}_{new}\equiv x_{new}$ and $\hat{d}_{new}\equiv
d_{new}.$

\textit{Proof.} Subtracting (\ref{equ1_equivalent_est}) from
(\ref{equ1_equivalent}) results in $\dot{\tilde{x}}_{new}=A\tilde{x}%
_{new},\tilde{x}_{new}\left(  0\right)  =0,$ where $\tilde{x}_{new}%
=x_{new}-\hat{x}_{new}.$ Then $\tilde{x}_{new}\equiv0$. This implies
that
$\hat{x}_{new}\equiv x_{new}.$ Consequently, by the relation $y=c^{T}%
x_{new}+d_{new}$ in (\ref{equ1_equivalent}), we have
$\hat{d}_{new}\equiv d_{new}.$ $\square$

\textbf{Remark 4}\textit{. }By\textit{\
}(\ref{equ1_equivalent_pro}), if the new state $x_{new}$ is bounded,
then the original state $x$ is bounded as well since the matrix $A$
is stable and the disturbance $d$ is bounded. This explains why the
matrix $A$ is chosen\ to be stable. To eliminate the transient
effect of initial values in $d_{new},$ namely $c^{T}e^{At}x_{0},$ we
often assign\ the eigenvalues for $A$ to have large negative real
part. By using the new state $x_{new}$, the controller can be
designed based on the transformed system (\ref{equ1_equivalent})
directly as shown in the following sections.

\textbf{Remark 5}\textit{. }It is interesting to note that the new
state\textit{\ }$x_{new}$ and\textit{\ }disturbance
$d_{new}$\textit{\ }in the transformed system
(\ref{equ1_equivalent}) can be observed directly rather than
asymptotically or exponentially. This will facilitate the analysis
and design later.\textit{\ }In practice, the output $y$ will be more
or less
subject to noise. In this case, the\textit{\ }stable\textit{\ }%
matrix\textit{\ }$A\ $will result in a small $\tilde{x}_{new}$ in
the presence of small noise, i.e. $\hat{x}_{new}\ $close to
$x_{new}$.

\textbf{Example 1}. A single-link robot arm with a revolute elastic
joint rotating in a vertical plane is served as an application in
this paper
\cite{Marino(1995)}:%
\begin{align}
\dot{x}_{1}  &  =x_{2}\nonumber\\
\dot{x}_{2}  &  =-\frac{F_{l}}{J_{l}}x_{2}-\frac{k}{J_{l}}\left(  x_{1}%
-x_{3}\right)  -\frac{Mgl}{J_{l}}\sin x_{1}+d_{1}\nonumber\\
\dot{x}_{3}  &  =x_{4}\nonumber\\
\dot{x}_{4}  &  =-\frac{F_{m}}{J_{m}}x_{4}+\frac{k}{J_{m}}\left(  x_{1}%
-x_{3}\right)  +\frac{1}{J_{m}}\tau+d_{2}\nonumber\\
y  &  =x_{1} \label{RTACmodel0}%
\end{align}
where $x_{1},x_{2},x_{3},x_{4}$ are the link displacement (rad),
link velocity (rad/s), rotor displacement (rad) and rotor velocity
(rad/s), respectively; $d_{1}$ and $d_{2}$ are unknown disturbance.
The initial value is assumed to be $x\left(  0\right)  =\left[
\begin{array}
[c]{cccc}%
0.05 & 0 & 0.05 & 0
\end{array}
\right]  ^{T}.$ Let link inertia $J_{l}=2$kg$\cdot$m$^{2},$ the
motor rotor inertia $J_{m}=0.5$kg$\cdot$m$^{2},$ the elastic
constant $k=0.05$kg$\cdot $m$^{2}$/s$,$ the link mass $M=0.5$kg, the
gravity constant $g=9.8$m/s$^{2}$,
the center of mass $l=0.5$m and viscous friction coefficients $F_{l}%
=F_{m}=0.2$kg$\cdot$m$^{2}$/s. The control $\tau$ is the torque
delivered by the motor. The control problem here is: assuming only
$y$ is measured, $\tau$
is to be designed so that $y$ tracks a smooth enough reference $r\ $%
asymptotically or with good tracking accuracy. The controller $\tau$
in (\ref{RTACmodel0})\ is designed as follows $\tau=J_{m}u,$ where
$u$ will be specified later. Then the system (\ref{RTACmodel0}) can
be written in form of
(\ref{equ0}) with%
\begin{equation}
A_{0}=\left[
\begin{array}
[c]{cccc}%
0 & 1 & 0 & 0\\
-\frac{k}{J_{l}} & -\frac{F_{l}}{J_{l}} & \frac{k}{J_{l}} & 0\\
0 & 0 & 0 & 1\\
\frac{k}{J_{m}} & 0 & -\frac{k}{J_{m}} & -\frac{F_{m}}{J_{m}}%
\end{array}
\right]  ,b=\left[
\begin{array}
[c]{c}%
0\\
0\\
0\\
1
\end{array}
\right]  ,c=\left[
\begin{array}
[c]{c}%
1\\
0\\
0\\
0
\end{array}
\right]  ,\phi_{0}\left(  y\right)  =\left[
\begin{array}
[c]{c}%
0\\
-\frac{Mgl}{J_{l}}\sin y\\
0\\
0
\end{array}
\right]  . \label{par}%
\end{equation}
It is easy to verify that the pair $\left(  A_{0},c\right)  $ is
observable. So for this application \textit{Assumption 1} holds. It
is found that $A_{0}$ is unstable. Choosing $p=\left[
\begin{array}
[c]{cccc}%
-2.10 & -1.295 & -9.36 & 3.044
\end{array}
\right]  ^{T},$ the system (\ref{RTACmodel0}) is formulated into
(\ref{equ1_equivalent}) with%
\[
A=\left[
\begin{array}
[c]{cccc}%
-2.1 & 1 & 0 & 0\\
-1.32 & -0.1 & 0.025 & 0\\
-9.36 & 0 & 0 & 1\\
3.144 & 0 & -0.1 & -0.4
\end{array}
\right]  ,\phi\left(  y\right)  =\left[
\begin{array}
[c]{c}%
2.1y\\
1.295y-1.225\sin y\\
9.36y\\
-3.044y
\end{array}
\right]  .
\]
where the eigenvalues of $A$ are assigned as $-0.5,$ $-0.6,$ $-0.7,$
$-0.8.$

\section{Additive State Decomposition of Transformed System}

In this section, the transformed system (\ref{equ1_equivalent}) is
`additively' decomposed into three independent subsystems in charge
of corresponding subtasks, namely the tracking subtask, the
rejection subtask and the stabilization subtask, as shown in Fig.1.
There exist many tools to analyze LTI systems, such as Laplace
transformation and transfer function, and state-space techniques.
Based on the above consideration, the transformed system
(\ref{equ1_equivalent}) is expected to be decomposed into two
subsystems by the additive state decomposition: an LTI system
including all external signals as the primary system, together with
the secondary system free of external signals. Therefore, the
original tracking task for the system (\ref{equ1_equivalent}) is
correspondingly decomposed into two subtasks by the additive state
decomposition: a tracking (including rejection) subtask for an LTI
`primary' system and a stabilization subtask for the left
`secondary' system. Since the tracking (including rejection) subtask
is only assigned to the LTI system, it therefore is a lot easier
than that for a nonlinear one. Furthermore, the tracking (including
rejection) subtask is decomposed into a pure tracking subtask and a
pure rejection subtask.\begin{figure}[h]
\begin{center}
\includegraphics[
scale=0.75 ]{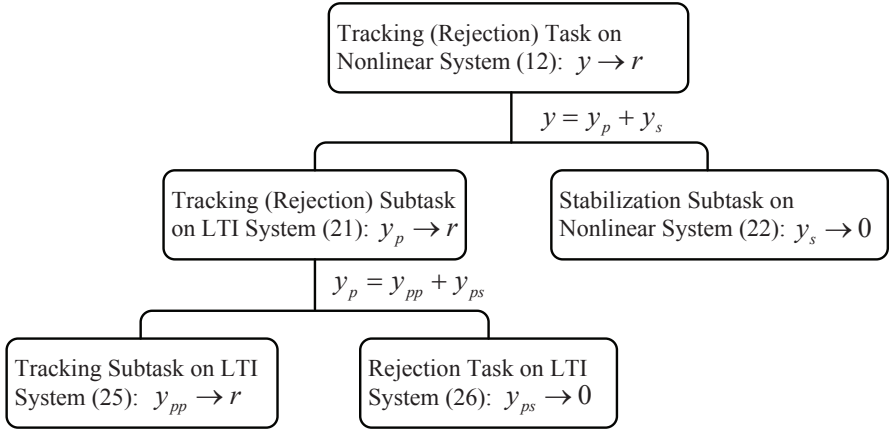}
\end{center}
\caption{Additive state decomposition flow}%
\end{figure}

Consider the transformed system (\ref{equ1_equivalent}) as the
original system. According to the principle above, we choose the
primary system as
follows:%
\begin{align}
\dot{x}_{p}  &  =Ax_{p}+bu_{p}+\phi\left(  r\right) \nonumber\\
y_{p}  &  =c^{T}x_{p}+d_{new},x_{p}\left(  0\right)  =0. \label{equ1_Pri}%
\end{align}
Then the secondary system is determined by the original system
(\ref{equ1_equivalent}) and the primary system (\ref{equ1_Pri}) with
the rule
(\ref{Dif_Sec_Sys}), and we can obtain that%
\begin{align}
\dot{x}_{s}  &  =Ax_{s}+bu_{s}+\phi\left(  r+e_{p}+c^{T}x_{s}\right)
-\phi\left(  r\right) \nonumber\\
y_{s}  &  =c^{T}x_{s},x_{s}\left(  0\right)  =0 \label{equ1_Sec}%
\end{align}
where $u_{s}=u-u_{p}$ and $e_{p}=y_{p}-r.$ According to (\ref{Gen_RelationDif}%
), we have%
\begin{equation}
x_{new}=x_{p}+x_{s}\text{ and }y=y_{p}+y_{s}. \label{equ1_relation}%
\end{equation}

The strategy here is to assign the tracking (including rejection)
subtask to the primary system (\ref{equ1_Pri}) and the stabilization
subtask to the
secondary system (\ref{equ1_Sec}). It is clear from (\ref{equ1_Pri}%
)-(\ref{equ1_relation}) that if the controller $u_{p}$ drives $y_{p}%
\rightarrow r$ and the controller $u_{s}$ drives $y_{s}\rightarrow0$
as $t\rightarrow\infty,$ then $y\rightarrow r$ as
$t\rightarrow\infty$. The benefit brought by the additive state
decomposition is that the controller $u_{s}$ will not affect the
tracking and rejection performance since the primary system
(\ref{equ1_Pri}) is independent of the secondary system
(\ref{equ1_Sec}). On the other hand, if the secondary system
(\ref{equ1_Sec}) is input-to-state stable with respect to the signal
$e_{p}$, then the dynamics of controller $u_{p}$ will not change the
input-to-state stability property. Therefore, conflict between
performance (trajectory tracking and disturbance rejection) and
robustness is avoided. Since the states $x_{p}$ and $x_{s}$ are
unknown except for sum of them, namely $x_{new}$, an observer is
proposed to estimate $x_{p}$ and $x_{s}.$

\textbf{Theorem 3}\textit{. }Under \textit{Assumption 1}, suppose
that an
observer is designed to estimate state $x_{p}$ and $x_{s}$ in (\ref{equ1_Pri}%
)-(\ref{equ1_Sec}) as follows:
\begin{subequations}
\label{equ1_Obs}%
\begin{align}
\hat{x}_{p}  &  =x_{new}-\hat{x}_{s}\label{equ1_Obs1}\\
\dot{\hat{x}}_{s}  &  =A\hat{x}_{s}+bu_{s}+\phi\left(  y\right)
-\phi\left(
r\right)  ,\hat{x}_{s}\left(  0\right)  =0. \label{equ1_Obs2}%
\end{align}
Then $\hat{x}_{p}\equiv x_{p}$ and $\hat{x}_{s}\equiv x_{s}.$

\textit{Proof. }Subtracting (\ref{equ1_Obs2}) from (\ref{equ1_Sec})
results in $\dot{\tilde{x}}_{s}=A\tilde{x}_{s},\tilde{x}_{s}\left(
0\right)  =0$ \footnote{Since the initial values $x_{new}\left(
0\right)  ,x_{s}\left( 0\right)  ,\hat{x}_{s}\left(  0\right)  $ are
all assigned by the designer, they are all determinate.}$,$ where
$\tilde{x}_{s}=x_{s}-\hat{x}_{s}.$ Then $\tilde{x}_{s}\equiv0$. This
implies that $\hat{x}_{s}\equiv x_{s}.$ Consequently, by
(\ref{equ1_relation}), we have $\hat{x}_{p}=x_{new}-\hat
{x}_{s}=x_{p}.$ $\square$

To avoid conflict between tracking performance and rejection
performance, the tracking subtask for the primary system
(\ref{equ1_Pri}) is further decomposed into two subtasks by the
additive state decomposition: a pure tracking subtask and a pure
rejection subtask. At this time, consider the primary system
(\ref{equ1_Pri}) as the original system and choose the primary
system of (\ref{equ1_Pri}) as follows:
\end{subequations}
\begin{align}
\dot{x}_{pp}  &  =Ax_{pp}+bu_{pp}+\phi\left(  r\right)  ,\nonumber\\
y_{pp}  &  =c^{T}x_{pp},x_{pp}\left(  0\right)  =0. \label{equ1_Pri_Pri}%
\end{align}
Then the secondary system of (\ref{equ1_Pri}) is determined by the
system (\ref{equ1_Pri}) and (\ref{equ1_Pri_Pri}) with the rule
(\ref{Dif_Sec_Sys})
that%
\begin{align}
\dot{x}_{ps}  &  =Ax_{ps}+bu_{ps}\nonumber\\
y_{ps}  &  =c^{T}x_{ps}+d_{new},x_{ps}\left(  0\right)  =0,
\label{equ1_Pri_Sec}%
\end{align}
where $u_{ps}=u_{p}-u_{pp}.$ According to (\ref{Gen_RelationDif}), we have%
\begin{equation}
x_{p}=x_{pp}+x_{ps}\ \text{and}\ y_{p}=y_{pp}+y_{ps}. \label{equ1_Pri_rela}%
\end{equation}
It is clear from (\ref{equ1_Pri_Pri})-(\ref{equ1_Pri_rela}) that if
the controller $u_{pp}$ drives $y_{pp}\rightarrow r$ and the
controller $u_{ps}$ drives $y_{ps}\rightarrow0$ as
$t\rightarrow\infty,$ then $y_{p}\rightarrow r$ as
$t\rightarrow\infty$. It is noticed that the controller $u_{pp}$ and
$u_{ps}$ above are independent each other. So, conflict between
tracking performance and rejection performance is avoided.

\section{Controller Design}

So far, we have transformed the original system to a
disturbance-free system whose state can be estimated directly. And
then, decompose the transformed system into three independent
subsystems in charge of corresponding subtasks. In this section, we
are going to investigate the controller design with respect to the
three decomposed subtasks respectively.

\subsection{Problem for Tracking Subtask}

\textbf{Problem 1}. \textit{For (\ref{equ1_Pri_Pri}), design a control input}%
\begin{align}
\dot{z}^{r}  &  =\alpha^{r}\left(  z^{r},x_{pp},r\right) \nonumber\\
u_{pp}  &  =u^{r}\left(  z^{r},x_{pp},r\right)  \label{Assumption2_Con}%
\end{align}
\textit{such that }$y_{pp}\rightarrow r$\textit{\ as
}$t\rightarrow\infty $\textit{.}

\textbf{Remark 6}\textit{\ (on Problem 1)}. \textit{Problem 1 }can
be
considered as a stable inversion problem \cite{Devasia(1996)}%
,\cite{Hunt(1997)}, which has been solved for the system in the form
of (\ref{equ1_Pri_Pri}) no matter whether it is minimum phase or
nonminimum phase. If the system (\ref{equ1_Pri_Pri}) is only minimum
phase, then the stable inversion problem is made a lot easier by
using the transfer function method. \textit{Problem 1 }can also be
considered as an output regulation problem
\cite{Ding(2003)}-\cite{Riccardo(2008)}, if the reference $r$ is
generated by an autonomous system. Since the reference $r$ is given,
the computation above can be done offline. In addition, since the
control design is only for a classical LTI system, the computation
is easier compared with that for a nonlinear one.

\textbf{Example 2}\textit{\ (Example 1 Continued)}. Since system
(\ref{RTACmodel0}) is minimum phase, by the transfer function
method, the
reference control input $u_{pp}=u^{r}$ can be designed as follows:%
\begin{equation}
u^{r}\left(  s\right)  =\frac{1}{c^{T}F^{-1}\left(  s\right)
b}\left[ r\left(  s\right)  +c^{T}F^{-1}\left(  s\right)  \left(
pr\left(  s\right) +b_{1}\frac{Mgl}{J_{l}}\left(  \sin r\right)
\left(  s\right)  \right)
\right]  \label{Assumption2_Sys_con}%
\end{equation}
where $u^{r}\left(  s\right)  $ is the transfer function of signal
$u^{r},$ $F\left(  s\right)  =sI_{4}-A$ and $b_{1}=\left(
\begin{array}
[c]{cccc}%
0 & 1 & 0 & 0
\end{array}
\right)  ^{T}$. Then $y_{pp}\rightarrow r$ as $t\rightarrow\infty.$
\textit{Problem 1} is solved.

\subsection{Problem for Rejection Subtask}

\textbf{Problem 2}. \textit{For (\ref{equ1_Pri_Sec}), there exists a
control
input}%
\begin{align}
\dot{z}^{d}  &  =\alpha^{d}\left(  z^{d},x_{ps},d_{new}\right) \nonumber\\
u_{ps}  &  =u^{d}\left(  z^{d},x_{ps},d_{new}\right)  \label{Assumption3_Con}%
\end{align}
\textit{such that }$y_{ps}\rightarrow\mathcal{B}\left(
\delta\right)  $
\footnote{$\mathcal{B}\left(  \delta\right)  \triangleq\left\{  \xi\in%
\mathbb{R}
\left\vert \left\Vert \xi\right\Vert \leq\delta\right.  \right\}  ,$
where $\delta=\delta\left(  d\right)  $ is a function of the
disturbance $d;$ the notation $x\left(  t\right)
\rightarrow\mathcal{B}\left(  \delta\right)  $ means
$\underset{y\in\mathcal{B}\left(  \delta\right)  }{\min}$
$\left\vert x\left(  t\right)  -y\right\vert
\rightarrow0.$}\textit{\ as }$t\rightarrow
\infty$\textit{. In particular, if }$\delta=0,$\textit{\ then }$y_{ps}%
\rightarrow0$\textit{\ as }$t\rightarrow\infty$\textit{.}

\textbf{Remark 7}\textit{\ (on Problem 2)}. Since the system
(\ref{equ1_Pri_Sec}) is a classical LTI system, some standard
designs in frequency domain, such as the transfer function method,
can be used to handle a general disturbance \cite{Morari(1989)}. In
this case, the disturbance cannot be often rejected asymptotically.
So, the \textit{Problem 2 }needs to\textit{\ }consider the result
$y_{ps}\rightarrow\mathcal{B}\left( \delta\right)  $ besides
$y_{ps}\rightarrow0.$ If $d_{new}$ is generated by an autonomous
system, then \textit{Problem 2 }can\textit{\ }be considered as an
output regulation problem \cite{Ding(2003)}-\cite{Bobtsov(2011)}. In
this case, the disturbance can be rejected asymptotically. The
technique in \cite{Ding(2003)}-\cite{Bobtsov(2011)} of course can be
still applied to the problem even if both parameters and order of an
exosystem are uncertain.

\textbf{Example 3}\textit{\ (Example 1 Continued)}. To demonstrate
the effectiveness of the proposed control, the disturbances in
(\ref{RTACmodel0}) are assumed to be in two cases. \textit{Case 1
(sinusoidal)}: the unknown disturbances $d_{1}$ and $d_{2}$ are
sinusoidal, such as $d_{1}=0.05\sin t\ $and $d_{2}=0.05\sin\left(
t+1\right)  .$ \textit{Case 2 (general)}: the unknown disturbances
$d_{1}$ and $d_{2}$ are driven by normal $\left(
\text{Gaussian}\right)  $ distributed random signals $\zeta_{1},\zeta_{2}%
\sim\mathcal{N}\left(  0,1\right)  .$ The transfer functions are
assumed to be $d_{1}\left(  s\right)  =\frac{0.08}{\left(
s+0.2\right)  \left( s+0.5\right)  \left(  s+0.8\right)
}\zeta_{1}\left(  s\right)  $ and $d_{2}\left(  s\right)
=\frac{6}{\left(  s+1\right)  \left(  s+2\right) \left(  s+3\right)
}\zeta_{2}\left(  s\right)  $. The resulting disturbances are shown
in Fig.2.

\textit{Case 1 (sinusoidal). }Since the unknown disturbances $d_{1}$
and
$d_{2}$ are sinusoidal with frequency $1$ rad/s, by (\ref{equ1_equivalent_pro}%
) we can conclude that $d_{new}=\bar{d}+\varepsilon$, where
$\bar{d}$ is sinusoidal with frequency $1$ rad/s as well and
$\varepsilon\rightarrow0$ as $t\rightarrow\infty.$ This implies
that$\ $the transfer function $\bar{d}$ can be written as
$\bar{d}\left(  s\right)  =\frac{1}{s^{2}+1}w\left(  0\right)
.$ For (\ref{equ1_Pri_Sec}), design controller $u_{ps}=u^{d}$ as follows%
\begin{align}
\dot{z}^{d}  &  =Sz^{d}+c_{d}y_{ps}\nonumber\\
u^{d}  &  =k_{1}^{T}z^{d}+k_{2}^{T}x_{ps} \label{Assumption3_Sys_con1}%
\end{align}
where $S=\left[
\begin{array}
[c]{cc}%
0 & 1\\
-1 & 0
\end{array}
\right]  ,$ $c_{d}=\left[
\begin{array}
[c]{c}%
1\\
0
\end{array}
\right]  ,$ $k_{1}=\left[
\begin{array}
[c]{cc}%
96 & 122
\end{array}
\right]  ^{T}$ and $k_{2}=\left[
\begin{array}
[c]{cccc}%
34 & 70 & -0.67 & -1.9
\end{array}
\right]  ^{T}.$ Combining (\ref{equ1_Pri_Sec}) with
(\ref{Assumption3_Sys_con1}) results in%
\begin{align*}
\dot{z}^{d}  &  =Sz^{d}+c_{d}c^{T}x_{ps}+c_{d}d_{new}\\
\dot{x}_{ps}  &  =k_{1}^{T}z^{d}+\left(  A+bk_{2}^{T}\right)  x_{ps}\\
y_{ps}  &  =c^{T}x_{ps}+d_{new}.
\end{align*}
The transfer function from $d_{new}$ to $y_{ps}$ possesses negative
real poles and at least two zeros $\pm i.$ Since $\bar{d}\left(
s\right)  =\frac {1}{s^{2}+1}w\left(  0\right)  $ and
$\varepsilon\rightarrow0,$ we have $y_{ps}\rightarrow0.$
\textit{Problem 2} is solved for \textit{Case 1}.

\textit{Case 2 (general)}.\textbf{\ }The transfer function from
$u_{ps}$ to
$y_{ps}$ can be represented as follows:%
\begin{equation}
y_{ps}\left(  s\right)  =c^{T}F^{-1}\left(  s\right)  bu_{ps}\left(
s\right) +d_{new}\left(  s\right)  .\nonumber
\end{equation}
Since $d_{new}$ is a low-frequency disturbance and can be observed
by (\ref{equ1_equivalent_est}), an easy way is to design the
disturbance
compensator as follows:%
\begin{equation}
u^{d}\left(  s\right)  =-\frac{1}{c^{T}F^{-1}\left(  s\right)
b}Q\left(
s\right)  d_{new}\left(  s\right)  \label{Assumption3_Sys_con2}%
\end{equation}
where $u^{d}\left(  s\right)  $ is the transfer function of $u^{d},$
$Q\left( s\right)  $ is a low-pass filter satisfying
\[
\left(  1-Q\left(  j\omega\right)  \right)  d_{new}\left(
j\omega\right)
\approx0,\forall\omega\in%
\mathbb{R}
.
\]
Moreover, $Q\left(  s\right)  $ is at least fourth order to make the
compensator physically realizable (the order of denominator is
greater than or
equal to that of numerator). In this simulation, we choose%
\[
Q\left(  s\right)  =\frac{1}{\prod\nolimits_{k=5}^{k=8}\left(  \frac{1}%
{10k}s+1\right)  }.
\]
In this case, $y_{ps}$ in (\ref{equ1_Pri_Sec}) is ultimately bounded
by a small value, namely $y_{ps}\rightarrow\mathcal{B}\left(
\delta\right)  $, where $\delta$ can be adjusted by $Q\left(
s\right)  .$ \textit{Problem 2} is solved for \textit{Case 2}.

\subsection{Problem for Stabilization Subtask}

\textbf{Problem 3}. \textit{For (\ref{equ1_Sec}), there exists a
controller }$u_{s}=u^{s}\left(  x_{s},r,\cdots,r^{\left(  N\right)
}\right) $\textit{\ such that the closed-loop system is
input-to-state stable with
respect to the input }$e_{p},$\textit{\ namely }%
\begin{equation}
\left\Vert x_{s}\left(  t\right)  \right\Vert \leq\beta\left(
\left\Vert x_{s}\left(  t_{0}\right)  \right\Vert ,t-t_{0}\right)
+\gamma\left( \underset{t_{0}\leq s\leq t}{\sup}\left\Vert
e_{p}\left(  s\right)
\right\Vert \right)  ,t\geq t_{0}, \label{Prob3}%
\end{equation}
\textit{\ \ where }$r^{\left(  N\right)  }$ denotes the $N$th
derivative of $r$, function $\beta$ is a class
$\mathcal{KL}$\textit{\ function and\ }$\gamma$\textit{\ is a class
}$\mathcal{K}$\textit{\ function \cite{Khalil(2002)}.}

\textbf{Remark 8}\textit{\ (on Problem 3)}. If $e_{p}$ is
nonvanishing, then \textit{Problem 3 }is a classical input-to-state
stability problem. Readers can refer to
\cite{Khalil(2002)},\cite{Sontag(2007)} for how to design a
controller satisfying input-to-state stability or how to prove the
designed controller satisfying input-to-state stability\textit{.} In
particular, if $e_{p}\rightarrow0$ as $t\rightarrow\infty,$ then
$x_{s}\rightarrow0$ as $t\rightarrow\infty$ by (\ref{Prob3}). In
addition, if $e_{p}\rightarrow0$ as $t\rightarrow\infty,$ then
input-to-state stability can be relaxed as well. In fact,
\textit{Problem 3 }only considers how\textit{\ }$x_{s}$ behaves as
$e_{p}\rightarrow0$ as $t\rightarrow\infty.$ The reference
\cite{Sussmann(1991)} discussed under what conditions the solution
of (\ref{equ1_Pri_Sec}) say $x_{s}^{e_{p}}$ and the solution of
(\ref{equ1_Pri_Sec}) with $e_{p}\equiv0\ $say $x_{s}^{\ast}$ satisfy
$\left\Vert x_{s}^{e_{p}}\left(  t\right)  -x_{s}^{\ast}\left(
t\right) \right\Vert \leq\theta e^{-\eta t},$ where $\theta,\eta>0.$
In this case, only stability of (\ref{equ1_Pri_Sec}) with
$e_{p}\equiv0$ needs to be considered rather than input-to-state
stability.

\textbf{Example 4}\textit{\ (Example 1 Continued)}. The system (\ref{equ1_Sec}%
) can be rewritten as%
\begin{align}
\dot{x}_{s,1}  &  =x_{s,2}\nonumber\\
\dot{x}_{s,2}  &  =-\frac{F_{l}}{J_{l}}x_{s,2}-\frac{k}{J_{l}}\left(
x_{s,1}-x_{s,3}\right)  -\frac{Mgl}{J_{l}}\left[  \sin\left(  r+x_{s,1}%
\right)  -\sin\left(  r\right)  \right]  +d_{e_{p}}\nonumber\\
\dot{x}_{s,3}  &  =x_{s,4}\nonumber\\
\dot{x}_{s,4}  &  =-\frac{F_{m}}{J_{m}}x_{s,4}+\frac{k}{J_{m}}\left(
x_{s,1}-x_{s,3}\right)  +u_{s}\nonumber\\
y  &  =x_{s,1} \label{equ1_Sec_re}%
\end{align}
where $d_{e_{p}}=-\frac{Mgl}{J_{l}}\left[  \sin\left(
r+e_{p}+x_{s,1}\right) -\sin\left(  r+x_{s,1}\right)  \right]  .$ By
the feedback linearization
method, design $u^{s}$ as follows%
\begin{equation}
u^{s}\left(  x_{s},r,\dot{r},\ddot{r}\right)
=\mu_{1}+\frac{J_{l}}{k}\left(
v+\mu_{2}\right)  \label{Assumption4_Sys_con}%
\end{equation}
where%
\begin{align*}
v  &  =-7.5x_{s,1}-19x_{s,2}-17\eta_{3}-7\eta_{4}\\
\mu_{1}  &  =-\eta_{3}+\frac{k}{J_{m}}x_{s,1}-\frac{k}{J_{m}}x_{s,3}%
-\frac{F_{m}}{J_{m}}x_{s,4}\\
\mu_{2}  &  =\frac{F_{l}}{J_{l}}\eta_{4}+\frac{Mgl}{J_{l}}\left(
\eta _{3}+\ddot{r}\right)  \cos\left(  x_{s,1}+r\right)
-\frac{Mgl}{J_{l}}\left[ \left(  x_{s,2}+\dot{r}\right)
^{2}\sin\left(  x_{s,1}+r\right)  +\ddot
{r}\cos\left(  r\right)  -\dot{r}^{2}\sin\left(  r\right)  \right] \\
\eta_{3}  &  =-\frac{F_{l}}{J_{l}}x_{s,2}-\frac{k}{J_{l}}\left(
x_{s,1}-x_{s,3}\right)  -\frac{Mgl}{J_{l}}\left[  \sin\left(  x_{s,1}%
+r\right)  -\sin\left(  r\right)  \right] \\
\eta_{4}  &  =-\frac{F_{l}}{J_{l}}\eta_{3}-\frac{k}{J_{l}}\left(
x_{s,2}-x_{s,4}\right)  -\frac{Mgl}{J_{l}}\left[  \left(
x_{s,2}+\dot {r}\right)  \cos\left(  x_{s,1}+r\right)
-\dot{r}\cos\left(  r\right)
\right] \\
x_{s}  &  =\left[
\begin{array}
[c]{cccc}%
x_{s,1} & x_{s,2} & x_{s,3} & x_{s,4}%
\end{array}
\right]  ^{T}.
\end{align*}
Substituting (\ref{Assumption4_Sys_con}) into (\ref{equ1_Sec_re}) results in%
\[
\dot{x}_{s}^{\prime}=A^{\prime}x_{s}^{\prime}+d_{e_{p}}^{\prime}%
\]
where%
\[
x_{s}^{\prime}=\left[
\begin{array}
[c]{c}%
x_{s,1}\\
x_{s,2}\\
\eta_{3}\\
\eta_{4}%
\end{array}
\right]  ,A^{\prime}=\left[
\begin{array}
[c]{cccc}%
0 & 1 & 0 & 0\\
0 & 0 & 1 & 0\\
0 & 0 & 0 & 1\\
-7.5 & -19 & -17 & -7
\end{array}
\right]  ,d_{e_{p}}^{\prime}=\left[
\begin{array}
[c]{c}%
0\\
d_{e_{p}}\\
0\\
0
\end{array}
\right]  .
\]
Since the matrix $A^{\prime}$ is stable and $\left\Vert
d_{e_{p}}^{\prime }\right\Vert \leq\frac{Mgl}{2J_{l}}\left\Vert
e_{p}\right\Vert $, there exist a class $\mathcal{KL}$\textit{\
}function $\beta^{\prime}$ and\textit{\ }a class $\mathcal{K}$\
function $\gamma^{\prime}$\textit{ }such that\textit{
}\cite{Khalil(2002)}%
\begin{equation}
\left\Vert x_{s}^{\prime}\left(  t\right)  \right\Vert
\leq\beta^{\prime }\left(  \left\Vert x_{s}^{\prime}\left(
t_{0}\right)  \right\Vert ,t-t_{0}\right)  +\gamma^{\prime}\left(
\underset{t_{0}\leq s\leq t}{\sup }\left\Vert e_{p}\left(  s\right)
\right\Vert \right)  ,t\geq t_{0}.
\end{equation}
Furthermore, by the definition of $x_{s}^{\prime},$ the
\textit{Problem 3} is solved.

\subsection{Controller Integration}

With the solutions of the three problems in hand, we can state

\textbf{Theorem 4}. Under \textit{Assumption 1},\textit{\ }suppose
i) \textit{Problems 1-3} are solved; ii) the controller for system
(\ref{equ0}) (or (\ref{equ1_equivalent})) is designed as

Observer:%
\begin{align}
\dot{\hat{x}}_{new}  &  =A\hat{x}_{new}+bu+\phi\left(  y\right)
,\hat
{x}_{new}\left(  0\right)  =0\nonumber\\
\dot{\hat{x}}_{s}  &  =A\hat{x}_{s}+bu^{s}+\phi\left(  y\right)
-\phi\left(
r\right)  ,\hat{x}_{s}\left(  0\right)  =0.\nonumber\\
\dot{\hat{x}}_{pp}  &  =A\hat{x}_{pp}+bu^{r}+\phi\left(  r\right)
,\hat
{x}_{pp}\left(  0\right)  =0\nonumber\\
\hat{d}_{new}  &  =y-c^{T}\hat{x}_{new},\hat{x}_{p}=\hat{x}_{new}-\hat{x}%
_{s},\hat{x}_{ps}=\hat{x}_{p}-\hat{x}_{pp} \label{mainobsever1}%
\end{align}

Controller:%
\begin{align}
\dot{z}^{r}  &  =\alpha^{r}\left(  z^{r},\hat{x}_{pp},r\right)
,z^{r}\left(
0\right)  =0\nonumber\\
\dot{z}^{d}  &  =\alpha^{d}\left(
z^{d},\hat{x}_{ps},\hat{d}_{new}\right)
,z^{d}\left(  0\right)  =0\nonumber\\
u_{p}  &  =u^{r}\left(  z^{r},\hat{x}_{pp},r\right)  +u^{d}\left(  z^{d}%
,\hat{x}_{ps},\hat{d}_{new}\right) \nonumber\\
u  &  =u_{p}+u^{s}\left(  \hat{x}_{s},r,\cdots r^{\left(  N\right)
}\right)
. \label{maincontroller1}%
\end{align}
Then the output of system (\ref{equ0}) (or (\ref{equ1_equivalent}))
satisfies that $y\rightarrow r+\mathcal{B}\left(  \delta+\left\Vert
c\right\Vert \gamma\left(  \delta\right)  \right)  $ as
$t\rightarrow\infty$. In particular, if $\delta=0,$ then the output
in system (\ref{equ0}) (or (\ref{equ1_equivalent})) satisfies that
$y\rightarrow r$ as $t\rightarrow \infty.$

\textit{Proof. See Appendix. }$\square$

\textbf{Remark 9}. The controllers $u_{pp}$ and $u_{ps}$ are
designed based on the LTI systems, to which both design methods in
frequency domain and time
domain can be applied. The controller (\ref{mainobsever1}%
)-(\ref{maincontroller1}) has both following salient features: i)
three degrees of freedom offered by three independent subcontrollers
$u^{r},u^{d}$ and $u^{s}.$ This is similar to the idea of
two-degree-of-freedom control \cite{Morari(1989)}. If $r$ and
$d_{new}$ behave differently, then controller can be chosen both for
good tracking of reference $r$ and good rejection of disturbance
$d_{new}.$ In addition, as the reference and/or disturbance change,
only the corresponding subcontroller needs to be modified rather
than the whole one. ii) The control signal $u^{r}\ $and$\ u^{d}\
$driven by reference and disturbance, considered as feedforward,
will not effect on the stability of the closed-loop system. iii) The
controller can deal with more general reference and disturbance
signals more easily because subcontrollers $u^{r}\ $and$\ u^{d}$ are
designed based on a simple LTI system\textit{.}

\textbf{Example 5}\textit{\ (Examples 1-4 Continued).} According to
(\ref{maincontroller1}) we design the controller as follows:%
\[%
\begin{array}
[c]{ll}%
\text{\textit{Case 1(sinusoidal)}:} & \left\{
\begin{array}
[c]{l}%
\dot{z}^{d}=Sz^{d}+c_{d}\left(  c^{T}\hat{x}_{ps}+\hat{d}_{new}\right) \\
u_{p}=u^{r}+\left(  k_{1}^{T}z^{d}+k_{2}^{T}\hat{x}_{ps}\right) \\
u=u_{p}+u^{s}%
\end{array}
\right. \\
\text{\textit{Case 2(general)}:} & \left\{
\begin{array}
[c]{l}%
u_{p}=u^{r}+u^{d}\\
u=u_{p}+u^{s}%
\end{array}
\right.
\end{array}
\]
where $\hat{x}_{p},\hat{x}_{s},\hat{x}_{pp},\hat{d}_{new}$ are given
by (\ref{mainobsever1}) with the parameters in (\ref{par}), $u^{r}$
is given by (\ref{Assumption2_Sys_con}), $u^{s}$ is given by
(\ref{Assumption4_Sys_con}) and the disturbance compensator $u^{d}$
in \textit{Case 1} is designed according to
(\ref{Assumption3_Sys_con1}), while in \textit{Case 2} the
disturbance compensator $u^{d}$ is designed according to
(\ref{Assumption3_Sys_con2}).

First, let us see the control performance of the primary system
(\ref{equ1_Pri}) and the secondary system (\ref{equ1_Sec}) in
\textit{Case 1 (sinusoidal disturbance)}. The evolutions of their
outputs are shown in Fig.3. As shown, it can be seen that
$y_{p}=y_{pp}+y_{ps}\rightarrow r$ as $t\rightarrow\infty$ and
$y_{s}\rightarrow0$ as $t\rightarrow\infty$. Therefore, by the
additive state decomposition, $y=y_{p}+y_{s}\rightarrow r\ $as
$t\rightarrow\infty$. This is confirmed by the simulation shown in
Fig.4. Secondly, let us see the control performance of the primary
system (\ref{equ1_Pri}) and the secondary system (\ref{equ1_Sec}) in
\textit{Case 2 (general disturbance)}. The evolutions of their
outputs are shown in Fig.5. As shown, it can be seen that $y_{p}\
$tracks $r$ with good performance and $y_{s}$ is ultimately bounded
by a small value. Therefore, by the additive state decomposition,
$y=y_{p}+y_{s}\ $tracks $r$ with a good performance. This is
confirmed by the simulation in Fig.6.

\textbf{Remark 10}. By the additive-state-decomposition-based
tracking\emph{\ }control scheme, it is seen from the simulation that
the transfer function method is applied to the tracking controller
design, which increases flexibility of the design. By benefiting
from it, both a sinusoidal and a general disturbance can be handled
in the same framework, where only the subcontroller $u^{d}$ needs to
be modified rather than the whole one.

\section{Conclusions}

In this paper, the output feedback tracking problem for a class of
systems with measurable nonlinearities and unknown disturbances was
considered. Our main contribution lies in the presentation of a new
decomposition scheme, named additive state decomposition, which not
only simplifies the controller design but also increases flexibility
of the controller design.

The proposed additive-state-decomposition-based tracking control
scheme was adopted to solve the output feedback tracking problem.
First, the considered system was transformed to an input-output
equivalent disturbance-free system. Then, by the additive state
decomposition, the transformed system was decomposed into three
subsystems in charge of three independent subtasks respectively: an
LTI system in charge of a pure tracking subtask, another LTI system
in charge of a pure rejection subtask and a nonlinear system in
charge of a stabilization subtask. Based on the decomposition, the
subcontrollers corresponding to three subsystems were designed
separately, which increases the flexibility of design. To
demonstrate its effectiveness, the proposed
additive-state-decomposition-based tracking control was applied to
the output feedback tracking problem for a single-link robot arm
with a revolute elastic joint rotating in a vertical plane.

\section{Appendix: Proof of Theorem 4}

It is easy to follow the proof in \textit{Theorems 2-3 }that the\textit{\ }%
observer (\ref{mainobsever1}) will make
\begin{align}
\hat{x}_{new}  &  \equiv x_{new},\hat{d}_{new}\equiv
d_{new},\hat{x}_{p}\equiv
x_{p},\nonumber\\
\hat{x}_{s}  &  \equiv x_{s},\hat{x}_{pp}\equiv
x_{pp},\hat{x}_{ps}\equiv
x_{ps}. \label{mainobseverresult}%
\end{align}
The remainder proof is composed of two parts: i) for
(\ref{equ1_Pri}), the controller $u_{p}$ drives $y_{p}\rightarrow
r+\mathcal{B}\left( \delta\right)  $ as $t\rightarrow\infty$, and
ii) based on the result of i), for (\ref{equ1_Sec}), the controller
$u_{s}$ drives $y_{s}\rightarrow \mathcal{B}\left(  \left\Vert
c\right\Vert \gamma\left(  \delta\right) \right)  $ as
$t\rightarrow\infty.$ Then the controller $u=u_{p}+u_{s}$ drives
$y\rightarrow r+\mathcal{B}\left(  \delta+\left\Vert c\right\Vert
\gamma\left(  \delta\right)  \right)  $ as $t\rightarrow\infty\ $in
system (\ref{equ0}) (or (\ref{equ1_equivalent})).

i) Suppose that \textit{Problems 1-2} are solved. By\
(\ref{Assumption2_Con}) and\textit{\ }(\ref{mainobseverresult}), the
controller $u_{pp}$ is designed
as follows:%
\begin{align*}
\dot{z}^{r}  &  =\alpha^{r}\left(  z^{r},\hat{x}_{pp},r\right)
,z^{r}\left(
0\right)  =0\\
u_{pp}  &  =u^{r}\left(  z^{r},\hat{x}_{pp},r\right)
\end{align*}
which can drive $y_{pp}\rightarrow r$ as $t\rightarrow\infty$ in
(\ref{equ1_Pri_Pri}). By (\ref{Assumption3_Con}) and (\ref{mainobseverresult}%
), the controller $u_{ps}$ is designed as follows:%
\begin{align*}
\dot{z}^{d}  &  =\alpha^{d}\left(
z^{d},\hat{x}_{ps},\hat{d}_{new}\right)
,z^{d}\left(  0\right)  =0\\
u_{ps}  &  =u^{d}\left(  z^{d},\hat{x}_{ps},\hat{d}_{new}\right)
\end{align*}
which will drive $y_{ps}\rightarrow\mathcal{B}\left(  \delta\right)
$ as $t\rightarrow\infty$ in (\ref{equ1_Pri_Sec})$.$ Combining the
two controllers $u_{pp}$ and $u_{ps}$ above results in the
controller for the primary system
(\ref{equ1_Pri}):%
\begin{align}
\dot{z}^{r}  &  =\alpha^{r}\left(  z^{r},\hat{x}_{pp},r\right)
,z^{r}\left(
0\right)  =0\nonumber\\
\dot{z}^{d}  &  =\alpha^{d}\left(
z^{d},\hat{x}_{ps},\hat{d}_{new}\right)
,z^{d}\left(  0\right)  =0\nonumber\\
u_{p}  &  =u^{r}\left(  z^{r},\hat{x}_{pp},r\right)  +u^{d}\left(  z^{d}%
,\hat{x}_{ps},\hat{d}_{new}\right)  . \label{equ1_Pri_con}%
\end{align}
Therefore, by (\ref{equ1_Pri_rela}), the controller
(\ref{equ1_Pri_con}) can drive $y_{p}\rightarrow r+\mathcal{B}\left(
\delta\right)  $ as $t\rightarrow\infty$.

ii) Let us look at the secondary system (\ref{equ1_Sec}). Suppose
that \textit{Problems 3} is solved. By (\ref{mainobseverresult}),
the controller $u_{s}=u^{s}\left(  \hat{x}_{s},r,\cdots,r^{\left(
N\right)  }\right)  $ can
drive the output $y_{s}\ $such that%
\begin{align*}
\left\Vert y_{s}\left(  t\right)  \right\Vert  &  \leq\left\Vert
c\right\Vert
\left\Vert x_{s}\left(  t\right)  \right\Vert \\
&  \leq\left\Vert c\right\Vert \beta\left(  \left\Vert x_{s}\left(
t_{0}\right)  \right\Vert ,t-t_{0}\right)  +\left\Vert c\right\Vert
\gamma\left(  \underset{t_{0}\leq s\leq t}{\sup}\left\Vert
e_{p}\left( s\right)  \right\Vert \right)  ,t\geq0.
\end{align*}
Based on the result of i), we get $e_{p}\rightarrow\mathcal{B}\left(
\delta\right)  $ as $t\rightarrow\infty.$ This implies that
$\left\Vert e_{p}\left(  t\right)  \right\Vert
\leq\delta+\varepsilon$ when $t\geq
t_{0}+T_{1}.$ Then%
\begin{align*}
\left\Vert y_{s}\left(  t\right)  \right\Vert  &  \leq\left\Vert
c\right\Vert \beta\left(  \left\Vert x_{s}\left(  t_{0}+T_{1}\right)
\right\Vert ,t-t_{0}-T_{1}\right)  +\left\Vert c\right\Vert
\gamma\left(  \underset {t_{0}+T_{1}\leq s\leq t}{\sup}\left\Vert
e_{p}\left(  s\right)  \right\Vert
\right)  ,t\geq t_{0}+T_{1},\\
&  \leq\left\Vert c\right\Vert \beta\left(  \left\Vert x_{s}\left(
t_{0}+T_{1}\right)  \right\Vert ,t-t_{0}-T_{1}\right)  +\left\Vert
c\right\Vert \gamma\left(  \delta+\varepsilon\right)  ,t\geq
t_{0}+T_{1}.
\end{align*}
Since $\left\Vert c\right\Vert \beta\left(  \left\Vert x_{s}\left(
t_{0}+T_{1}\right)  \right\Vert ,t-t_{0}-T_{1}\right)  \rightarrow0$
as $t\rightarrow\infty\ $and $\varepsilon$ can be chosen arbitrarily
small, we can conclude $y_{s}\rightarrow\mathcal{B}\left(
\left\Vert c\right\Vert \gamma\left(  \delta\right)  \right)  $ as
$t\rightarrow\infty.$ Since $y=c^{T}x_{p}+c^{T}x_{s},$ we can
conclude that, driven by the controller (\ref{maincontroller1}), the
output of the system (\ref{equ0}) (or (\ref{equ1_equivalent}))
satisfies that $y\rightarrow r+\mathcal{B}\left( \delta+\left\Vert
c\right\Vert \gamma\left(  \delta\right)  \right)  $ as
$t\rightarrow\infty$. In particular, if $\delta=0,$ then the output
in system ((\ref{equ1_equivalent})) satisfies that $y\rightarrow r$
as $t\rightarrow \infty.$ $\square$

\begin{figure}[h]
\begin{center}
\includegraphics[
scale=0.65 ]{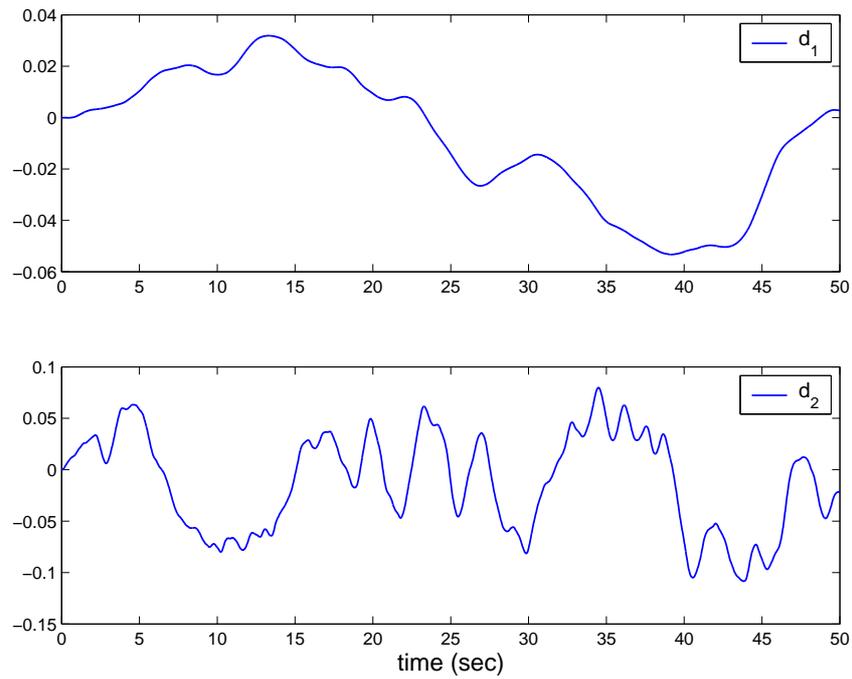}
\end{center}
\caption{Disturbance in \textit{Case 2}}%
\end{figure}\begin{figure}[ph]
\begin{center}
\includegraphics[
scale=0.65 ]{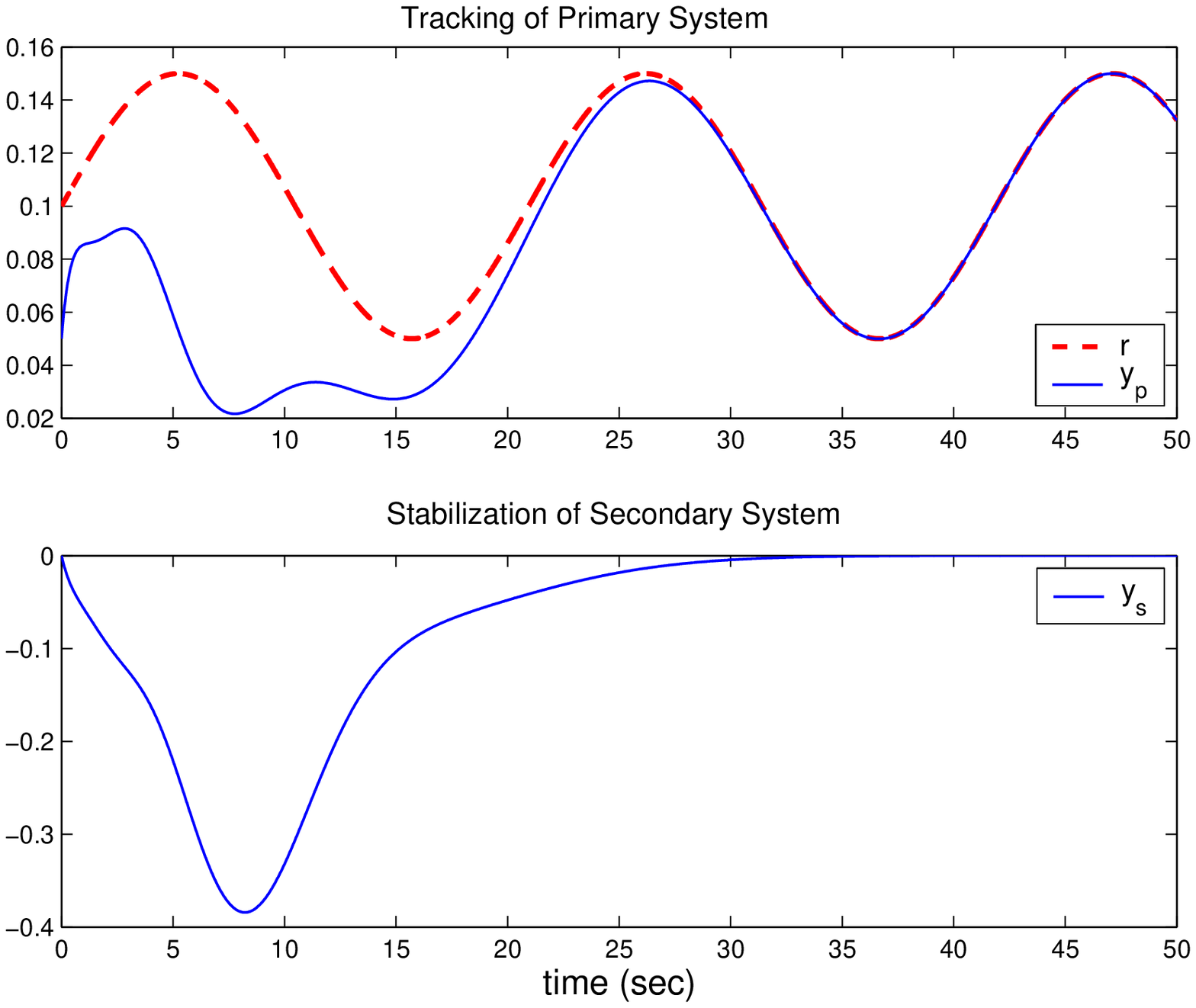}
\end{center}
\caption{Outputs of primary system and secondary system in \textit{Case 1}}%
\end{figure}\begin{figure}[ph]
\begin{center}
\includegraphics[
scale=0.65 ]{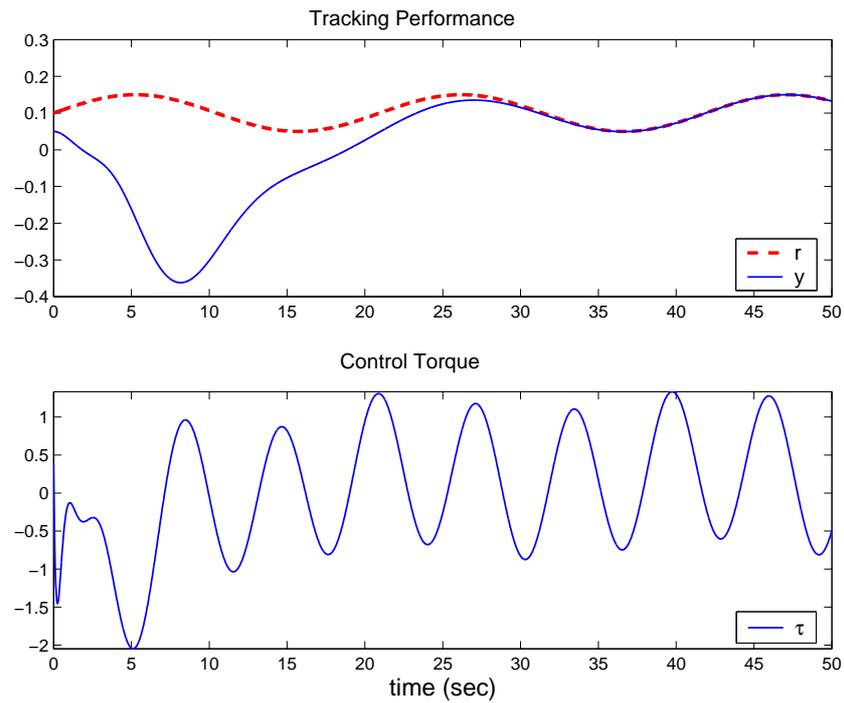}
\end{center}
\caption{Output (link displacement) of the single-link robot arm and
control
toque in \textit{Case 1}}%
\end{figure}\begin{figure}[h]
\begin{center}
\includegraphics[
scale=0.65 ]{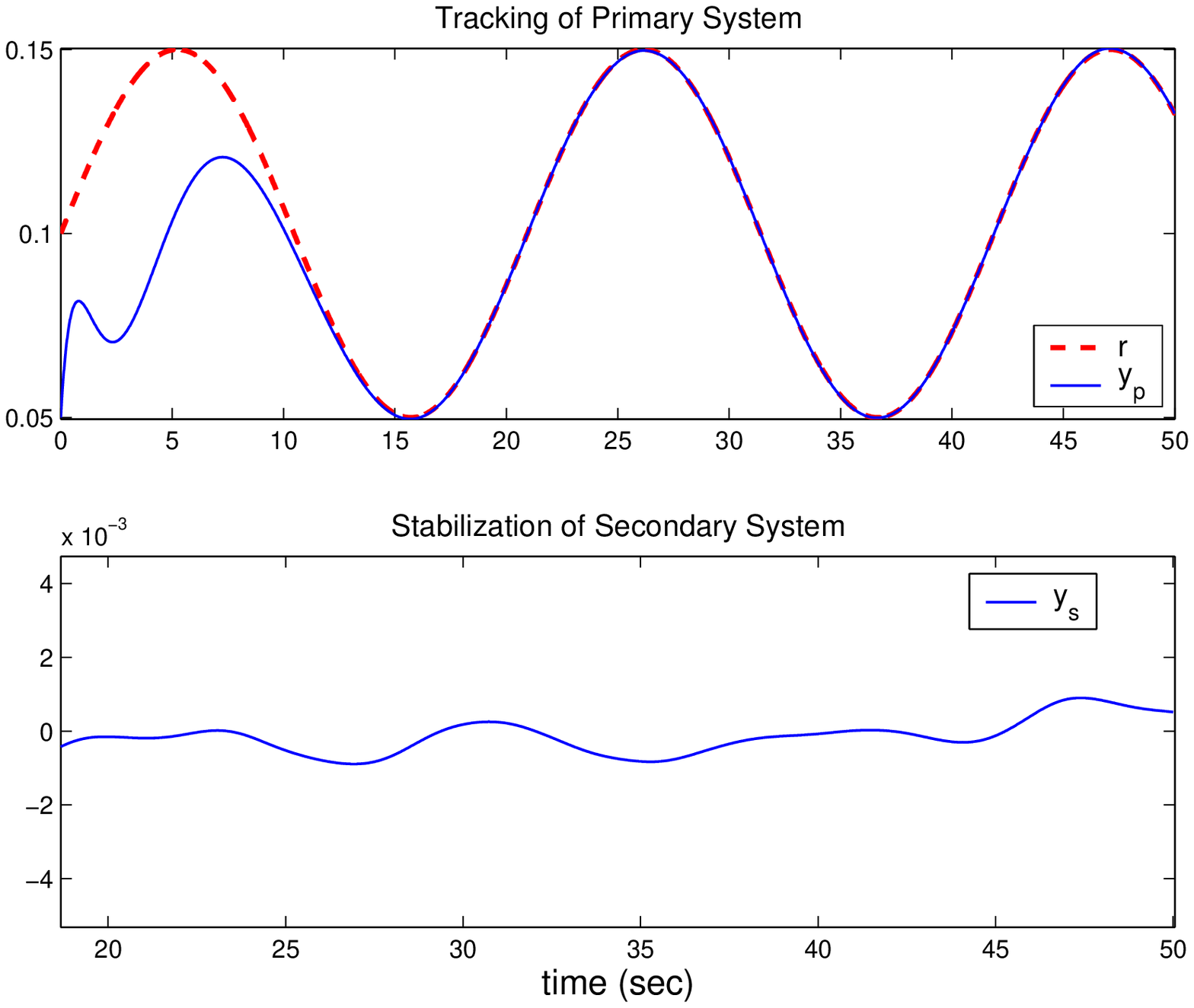}
\end{center}
\caption{Outputs of primary system and secondary system in \textit{Case 2}}%
\end{figure}\begin{figure}[h]
\begin{center}
\includegraphics[
scale=0.65 ]{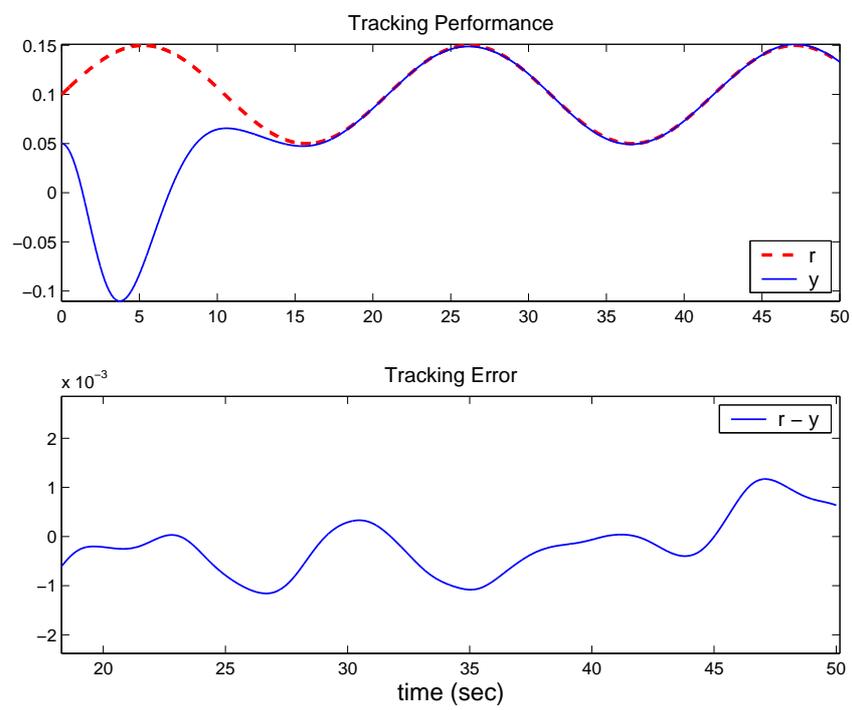}
\end{center}
\caption{Output (link displacement) of the single-link robot arm in
\textit{Case 2}}%
\end{figure}

\end{document}